\documentclass[%
 reprint,
 amsmath,amssymb,
 aps,
 showkeys,
]{revtex4-2}

\usepackage{graphicx}
\usepackage{dcolumn}
\usepackage{bm}

\begin{document}


\title{Taming Fluctuations for Gaussian States in Loop Quantum Cosmology}

\author{Patrick Fraser}
 \altaffiliation[Also at ]{Department of Philosophy, University of Toronto, Toronto, Ontario, Canada M5R 2M8}
 \email{p.fraser@mail.utoronto.ca}
\affiliation{%
 Department of Physics, University of Toronto, Toronto, Ontario, Canada M5S 1A7.
}%

\date{\today}

\begin{abstract}
We do not observe quantum effects on cosmological scales. Thus, if loop quantum cosmology (LQC) is to provide an accurate depiction of the real world, it must allow for quantum states of spacetime geometry which are semi-classical in two respects: they must be sharply peaked around a single, classical geometry, and they must have small quantum fluctuations. It is generally assumed that Gaussian states exhibit both of these properties. After all, they do in ordinary quantum mechanics. In this paper, we derive exact closed-form expressions for the fluctuations of Gaussian states in LQC and their lower bound given by the Robertson-Schr\"odinger inequality. We demonstrate that, contrary to ordinary quantum mechanics, fluctuations for Gaussian states in spatially flat, homogeneous and isotropic LQC diverge as the state variance increases (as well as in related cosmological models with the same kinematic Hilbert space and canonical observables). However, when the holonomy length is made to scale with a volume regularization parameter, these fluctuations may be arbitrarily suppressed by taking the fiducial volume to be large, providing analytic control over their divergence. Finally, we show that, despite this, Gaussian states in LQC generally do not minimize uncertainty. Moreover, it is conjectured that no such minimal-uncertainty states exist. Throughout this work, it becomes clear how important the often-assumed condition of holonomy length volume-scaling is; we show that when this condition is violated, the resulting theory exhibits operator closure pathologies and other exotic algebraic features.
\end{abstract}

\maketitle


\section{Introduction}
\label{sec:introduction}
In loop quantum cosmology (LQC), spacetime exists in a quantum state which may be regarded as a superposition of geometries. However, the observable universe appears to be sharp; one does not observe coherent quantum effects on cosmological scales. Thus, the \textit{actual} universe, as represented by LQC, must be described by a state which is sharply peaked around a single geometry. Moreover, such quantum geometries exhibit the same sort of quantum fluctuations as are present in ordinary quantum mechanics. Classical general relativity does not exhibit these fluctuations so the semi-classical states of LQC are expected to be states for which these fluctuations are very small, ideally vanishingly so.

These two properties -- sharpness and minimal fluctuation -- may be taken to be constitutive features of semi-classical states (see, for instance,~\cite{martin-dussaud:2020,corichi:2012,ashtekar:2005}). In ordinary quantum mechanics, such states are easy to find: Gaussian states -- and coherent states more generally -- satisfy both of these conditions. However, the Hilbert space structure of ordinary quantum mechanics is much different from that of LQC. Therefore, we cannot hastily assume that the same intuitions which hold for such `nice' states in ordinary quantum mechanics also hold for the analogous states in LQC.

In Section~\ref{sec:loop-quantum-cosmology} we introduce the basic theory of LQC. We then discuss general features of uncertainty relations in Section~\ref{sec:uncertainty}. In Section~\ref{sec:gaussian-states}, we define Gaussian states in the volume representation of LQC and explicitly compute the fluctuations of the canonical LQC observables for these Gaussian states in Section~\ref{sec:gaussian-fluctuations}. We show that there are two important cases to consider: that in which the shift operator is closed on the superselection sector (whence the holonomy length scales with a fiducial volume $V_0$) and otherwise. We show that the fluctuations of Gaussian states may diverge as their variance increases in either case for a fixed $V_0$, but that this divergence may be suppressed by taking $V_0\to\infty$ in the case where holonomy length is scales with $V_0$. In Section~\ref{sec:saturation}, we show that Gaussian states do not saturate uncertainty, and so this measure of semi-classicality fails for Gaussian states in general (though it is asymptotically satisfied when $V_0$ scaling is introduced). Finally, in Section~\ref{sec:squeezed-states} we sketch a further argument that there are in fact no physical squeezed states which minimize uncertainty.

\section{Loop Quantum Cosmology}
\label{sec:loop-quantum-cosmology}
LQC describes superpositions of classical spacetime geometries resulting in a theory of cosmology with several nice features~\cite{bojowald:2005,ashtekar:2011}, such as a bouncing evolution with no spacetime singularities~\cite{bojowald:2008,ashtekar:2006,wilson-ewing:2013,cai:2014}. As a simplifying assumption, we shall restrict ourselves to FLRW-spacetimes. That is, we assume spacetime is a superposition of geometries which each have a line element of the form

\begin{equation*}
    ds^2=-dt^2+a(t)^2[dx^2+dy^2+dz^2].
\end{equation*}

Such classical spacetimes may be foliated into time-like slices which each carry a fiducial 3-metric $\mathring{q}_{ab}$ (whose determinant is denoted $\mathring{q}$). For classical FLRW-spacetimes, we may encode this fiducial metric using Ashtekar variables given by

\begin{equation*}
    A_a^i=\dot{a}(t)(dx^i)_a,\qquad E_i^a=a(t)^2\sqrt{\mathring{q}}(\partial_i)^a.
\end{equation*}

Defining $c:=\dot{a}(t)$ and $p:=a(t)^2$, these become

\begin{equation*}
    A_a^i=c(dx^i)_a,\qquad E_i^a=p\sqrt{\mathring{q}}(\partial_i)^a.
\end{equation*}

The action which describes the dynamics of such spacetimes under the Einstein Field Equations is the Holst action, which has a symplectic term~\cite{rovelli:2014,rovelli:2015} of the form

\begin{equation*}
    \frac{1}{8\pi G\gamma}\int\dot{A}^i_aE^a_i d^3x=\frac{1}{8\pi G\gamma}\int\dot{c}p \sqrt{\mathring{q}}d^3x.
\end{equation*}

Unless the spacetime in question is compact, this term is divergent and must be regularized. Thus, one considers a spacetime region $\mathcal{V}$ with a finite fiducial volume $V_0=\int_\mathcal{V}\sqrt{\mathring{q}}$. Since FLRW-spacetimes are homogeneous, the location of this region is arbitrary. It is common (in both the classical and quantum theories) to scale the canonical coordinates with $V_0$ such that its dependency is transformed away~\cite{ashtekar:2011,ashtekar:2006,ashtekar:2003}. However, when we switch to the quantum theory, it has been shown that $V_0$ is not a gauge quantity~\cite{corichi:2012} (indeed, even after transforming it away from the canonical variables, the symplectic form still depends on it~\cite{ashtekar:2011}), and so we leave it for now. It shall play an interesting role in the analysis to come.

With this regularization, the Poisson bracket of $c$ and $p$ is given by

\begin{equation*}
    \{c,p\}=\frac{8\pi G\gamma}{V_0}
\end{equation*}

\noindent where $\gamma$ is the Barbero-Immirzi parameter. While this bracket is invariant under re-scaling, it still dependents on $V_0$~\cite{corichi:2012}. It is common now to make a canonical coordinate transformation, taking $V=p^{3/2}=a(t)^3$ to be the spatial volume of a time-like slice of the foliation at $t$, and taking $\beta=c/\sqrt{p}=\dot{a}(t)/a(t)$ to be the Hubble parameter (which is the conjugate momentum to the volume coordinate). The canonical variables $V$ and $\beta$ may then be readily use to study the cosmological behaviour of FLRW-spacetimes under classical Hamiltonian general relativity. These are the variables which are quantized to yield a quantum theory of cosmology.

In Wheeler-de Witt theory, one quantizes $V$ and $\beta$ using the usual Dirac quantization procedure. In LQC, we instead quantize the canonical variables using the quantization procedure of loop quantum gravity (LQG) -- which makes use of an alternative representation of the Weyl algebra -- to obtain a different quantum cosmological model. In this sense, LQC is essentially a symmetry reduced model of LQG with only a single degree of freedom (the scale factor $a(t)$ on each time slice).

Taking volume to be the coordinate observable of the resulting quantum theory of spatially flat, homogeneous and isotropic cosmology, the kinematic volume Hilbert space (in the so-called polymer representation~\cite{corichi:2007}) is the non-separable Hilbert space containing all \textit{countable} complex linear combinations of the basis $\{|V\rangle|V\in\mathbb{R}\}$ (whose coefficients $c_i$ satisfy $\sum_i|c_i|^2<\infty$). The inner-product is defined by $\langle V_x|V_y\rangle=\delta_{xy}$, the Kronecker-$\delta$. The sign of $V$ indicates manifold orientation. Thus, for any positive volume $V$, the value $-V$ here indicates the same volume with the opposite orientation, not a negative volume as such; there is no new physics introduced with negative values of $V$. (That said, it has been proposed by Christodoulou et al.~\cite{christodoulou:2012} that fermionic phases could be used to detect this orientation.)

There are two natural of operators to consider on this space; the volume operator $\hat{V}$, and the holonomy operator $\hat{h}_\lambda$ (also called the shift), which generates translations in the volume coordinate. Following the loop quantization procedure, these two operators are defined on the polymer Hilbert space in the following way:

\begin{equation}
    \hat{V}|V\rangle=V|V\rangle,\qquad\hat{h}_\lambda|V\rangle=\left|V+\frac{\alpha \lambda}{V_0}\right\rangle
\end{equation}

\noindent where $\alpha=4\pi G\gamma\hbar$ is a constant and $\lambda$ is the holonomy length. We see that $\hat{h}_\lambda$ depends on $V_0$. This will be important in what follows, for it ensures that the $\hat{h}_\lambda$ operator is never closed on superselection sectors unless $\lambda$ is made to scale with $V_0$.

While a coordinate translation arises from exponentiating the conjugate momentum in ordinary quantum mechanics, complications arise when one tries to extract a conjugate momentum observable $\hat{\beta}$ from $\hat{h}_\lambda$ (i.e. a holonomy \textit{flux} operator), where periodicity and other issues become apparent (see, for instance,~\cite{ashtekar:2003}). Specifically, on the polymer Hilbert space, $\hat{h}_\lambda$ is not strongly continuous in $\lambda$, and therefore the Stone-von Neumann theorem which usually ensures a unique representation of quantum mechanics fails. Thus, there does not exist a unique operator $\hat{\beta}$ for which $\hat{h}_{\lambda}=\exp\{-i\lambda\hat{\beta}\}$. A different choice of conjugate momentum may therefore be made, and it is here that the representation of LQG is different from that of ordinary quantum mechanics.

The canonical observables of LQC are $\hat{V}$ and $\frac{1}{\lambda}\hat{S}_\lambda$ where $\hat{S}_\lambda:=\frac{i}{2}(\hat{h}_\lambda-\hat{h}^*_\lambda)$. We may also define $\hat{C}_\lambda:=\frac{1}{2}(\hat{h}_\lambda+\hat{h}^*_\lambda)$.These operators are both self-adjoint and if there did exist a unique $\hat{\beta}$ such that $\hat{h}_\lambda$ were its exponentiation, then $\hat{S}_\lambda$ and $\hat{C}_\lambda$ would look like its sine and cosine and in small $\lambda$, $\frac{1}{\lambda}\hat{S}_\lambda$ would approximate $\hat{\beta}$, whence it resembles the conjugate momentum of $\hat{V}$.

These are the observables which enable one to make use of the Hamiltonian formalism to study the dynamics of quantum models of a spatially flat, homogeneous and isotropic spacetime. However, the analysis to come occurs solely in the kinematic Hilbert space in which one defines quantum states for spatial geometries in terms of their volume. That is, the dynamics are irrelevant for our considerations. As such, the following discussion is in principle applicable to a broader class of quantum cosmological models. Spatial homogeneity is needed for volume regularization to be well-motivated. Otherwise, any spacetime geometries whose quantization yields the above kinematic Hilbert space and which takes $\hat{V}$ and $\frac{1}{\lambda}\hat{S}_\lambda$ to be the canonical observables may be subjected to this analysis.

The most comprehensive summary of LQC beyond the spatially flat, homogeneous and isotropic case is found in~\cite{ashtekar:2011}. Here, we see that, while the Hamiltonian constraint which generates the dynamics of LQC is different for different models, the kinematic Hilbert space and canonical observables are unchanged in the positive-curvature ($k=+1$) FLRW-spacetimes as well as those models with a non-zero cosmological constant, so much of our analysis survives in these settings as well. The $\Lambda>0$ case yields challenges in ensuring that the phase space may be extended over the entire dynamics, but this does not impact the kinematics at a fixed time. Negative-curvature models ($k=-1$) require different operators and so demonstrating analogous results in that setting is non-trivial.

There is an important caveat when considering different cosmological models: the arbitrary scaling of the fiducial volume $V_0$ which plays an important role in gaining analytic control over divergent fluctuations is only possible in spatially non-compact spacetimes. In compact spacetimes (such as $k=+1$), one may only take $V_0$ to be the actual finite volume of spacetime and no larger. In such cases, the `taming' procedure described is of limited use.

With these basic notions from LQC established, we now discuss the fluctuations of these observables.

\section{Uncertainty}
\label{sec:uncertainty}
Let $\hat{A}$ and $\hat{B}$ be two symmetric operators on some Hilbert space $\mathcal{H}$. Then the Robertson-Schr\"odinger uncertainty relation is given by

\begin{equation}\label{eq:RS-Uncertainty}
    \Delta_A^2\Delta_B^2\geq \left|\frac{1}{2}\langle\{\hat{A},\hat{B}\}\rangle-\langle\hat{A}\rangle\langle\hat{B}\rangle\right|^2+\left|\frac{1}{2i}\langle[\hat{A},\hat{B}]\rangle\right|^2
\end{equation}

\noindent where $\{\cdot,\cdot\}$ is the anti-commutator~\cite{robertson:1929,schrodinger:1930}. The Robertson-Schr\"odinger relation is simply a more strict bound on the usual Heisenberg uncertainty relation. Indeed, the Heisenberg relation is obtained from Equation~\eqref{eq:RS-Uncertainty} by truncating the first term (which is always non-negative) yielding

\begin{equation}\label{eq:Heis-Uncertainty}
    \Delta_A^2\Delta_B^2\geq\left|\frac{1}{2i}\langle[\hat{A},\hat{B}]\rangle\right|^2.
\end{equation}

It is not \textit{a priori} obvious that there exist states which saturate the either~\eqref{eq:RS-Uncertainty} or~\eqref{eq:Heis-Uncertainty} for a given pair of observables, but saturation of the latter implies saturation of the former. The inequality~\eqref{eq:RS-Uncertainty} is well-defined provided it is evaluated for states $|\psi\rangle$ for which $\hat{A}|\psi\rangle$ is in the domain of $\hat{B}$ and vice-versa~\cite{davidson:1965,hall:2013}. We see that $\hat{V}$ and $\hat{h}_\lambda$ satisfy this condition for all LQC states.

Rovelli and Wilson-Ewing~\cite{rovelli:2014} provided a detailed analysis of the influence of fluctuations on cosmological effects by describing the large-scale phenomenology of LQC under the assumption that the lower bound of the Heisenberg relation is saturated. However, these results offer limited clarity without having on hand any states which saturate~\eqref{eq:Heis-Uncertainty}. The typical choice for such states from ordinary quantum mechanics are Gaussian states. We shall see, however, that these states do not minimize fluctuation generically in the LQC setting.

Let us now calculate the Robertson-Schr\"odinger fluctuations for these observables in the polymer representation. The commutator and anti-commutator of $\hat{V}$ with $\hat{h}_\lambda$ and $\hat{h}^*_\lambda$ are:

\begin{gather}\label{eq:comm-h}
    [\hat{V},\hat{h}_\lambda]=\frac{\alpha\lambda}{V_0}\hat{h}_\lambda,\qquad\{\hat{V},\hat{h}_\lambda\}=\hat{h}_\lambda\left(2\hat{V}+\frac{\alpha\lambda}{V_0}\hat{I}\right)\\\label{eq:comm-h*}
    [\hat{V},\hat{h}_\lambda^*]=-\frac{\alpha\lambda}{V_0}\hat{h}^*_\lambda,\qquad\{\hat{V},\hat{h}_\lambda^*\}=\hat{h}^*_\lambda\left(2\hat{V}-\frac{\alpha\lambda}{V_0}\hat{I}\right)
\end{gather}

Where $\hat{I}$ is the identity operator. Substituting~\eqref{eq:comm-h} and~\eqref{eq:comm-h*} into~\eqref{eq:RS-Uncertainty} yields

\begin{widetext}
\begin{equation}
    \Delta_V^2\Delta_{\frac{1}{\lambda}S_\lambda}^2\geq\left|\frac{i}{4\lambda}\left\langle\frac{\alpha\lambda}{V_0}(\hat{h}_\lambda-\hat{h}^*_\lambda)\right\rangle+\frac{i}{2\lambda}\langle(\hat{h}_\lambda-\hat{h}^*_\lambda)\hat{V}\rangle-\langle\hat{V}\rangle\left\langle\frac{i}{2\lambda}(\hat{h}_\lambda-\hat{h}^*_\lambda)\right\rangle\right|^2+\left|\frac{\alpha}{4V_0}\langle\hat{h}_\lambda+\hat{h}^*_\lambda\rangle\right|^2.
\end{equation}
\end{widetext}

But we may readily compute that $\langle(\hat{h}_\lambda-\hat{h}^*_\lambda)\hat{V}\rangle=\langle\hat{V}\rangle\langle\hat{h}_\lambda-\hat{h}^*_\lambda\rangle$. Thus, we have

\begin{equation}
    \Delta_V^2\Delta_{\frac{1}{\lambda}S_\lambda}^2\geq\left|\frac{i\alpha}{4V_0}\langle(\hat{h}_\lambda-\hat{h}^*_\lambda)\rangle\right|^2+\left|\frac{\alpha}{4V_0}\langle\hat{h}_\lambda+\hat{h}^*_\lambda\rangle\right|^2.
\end{equation}

This may be re-written as

\begin{equation}\label{eq:RS-volume-flux}
    \Delta_V^2\Delta_{\frac{1}{\lambda}S_\lambda}^2\geq\left(\frac{\alpha}{2V_0}\right)^2\left(|\langle\hat{S}_\lambda\rangle|^2+|\langle\hat{C}_\lambda\rangle|^2\right).
\end{equation}

When does this expression reduce to the Heisenberg uncertainty relation~\eqref{eq:Heis-Uncertainty}, i.e. when does $\langle\hat{S}_\lambda\rangle=0$? Without determining a solution in general, we note that any state $|\psi\rangle$ satisfying $\psi(V)=\psi(-V)$ will do this. Since spacetime is symmetric under a parity transformation (i.e. a change of orientation taking $V\mapsto -V$), this holds in general for physically meaningful LQC states.

Rovelli and Wilson-Ewing~\cite{rovelli:2014} discuss the behaviour of LQC states which are sharply peaked by analyzing the lower bound of the Heisenberg relation~\eqref{eq:Heis-Uncertainty} and supposing it is saturated by some such state. However, they do not explicitly construct any such states to show that this bound is saturated. It is this gap which we now fill in. It will be shown that, contrary to popular intuition, Gaussian state do \textit{not} saturate this lower bound. Nevertheless, under the right conditions, their fluctuations do approach the lower bound asymptotically in the limit considered by Rovelli and Wison-Ewing.

\section{Gaussian States}
\label{sec:gaussian-states}
In ordinary quantum mechanics, Gaussian states play a special role, for they are sharply peaked. Their probability amplitudes are centered on a particular point and decay exponentially away from that point at a rate over which we may have analytic control (by manipulating the variance $\sigma$). Additionally, they minimize the Heisenberg uncertainty relation~\eqref{eq:Heis-Uncertainty} between position and momentum observables. These two facts offer Gaussian states as a natural choice for semi-classical states: classical systems have no quantum fluctuations, and have definite locations; Gaussian states approximate both of these features to maximal precision. Sharpness is an obvious feature of Gaussian states. However, the fact that they minimize fluctuations is not trivial.

It has been argued~\cite{martin-dussaud:2020} that the \textit{constancy} of fluctuations of Gaussian states is also a crucial ingredient their semi-classicality. Time independence ensures that their fluctuations do not spread out under dynamical evolution, and thus \textit{remain} minimal. While this is certainly important, the discussion to follow shall show that even the condition of \textit{instantaneous} minimization is often too much to ask in the LQC setting (i.e. on a single time slice of the foliation). We therefore leave the study of the time evolution of these fluctuations open for future investigation.

In the volume representation of LQC, following Willis~\cite[Eq. 2.4.11]{willis:2004}, Gaussian states centered at $V=0$ take the form

\begin{equation}\label{eq:gaussian-states}
    |\psi\rangle=c\sum_{n\in\mathbb{Z}} e^{-(nl)^2/2\sigma^2}|nl\rangle
\end{equation}

\noindent where $l$ is a chosen lattice spacing and $\sigma>0$ is the Gaussian variance. The value $c$ is a normalization constant. An analogous definition for more general coherent states may be found in~\cite[p. 257]{ashtekar:2003}. In LQC, we choose $l$ to be minimal by taking it to be the Planck length $l_P$. The normalization is given by

\begin{equation}\label{eq:norm}
    |c|^2=\frac{1}{\sum_{n\in\mathbb{Z}}e^{-(nl)^2/\sigma^2}}.
\end{equation}

(We here normalize to 1.) If we wish to consider Gaussian states \textit{not} centered at $V=0$, we must make a small modification. Spacetime is thought to be have a parity symmetry under changes in manifold orientation. While field theories \textit{within} spacetimes may violate parity symmetries (e.g. the weak interaction in the Standard Model~\cite{wu:1957}), the underlying spacetime itself does not. Hence, we require that quantum states exhibit this parity symmetry as well, and thus satisfy $\psi(V)=\psi(-V)$~\cite{ashtekar:2011,bentivegna:2008}. However, Gaussian states as defined in~\eqref{eq:gaussian-states} violate this symmetry if one simply shifts the given state by a certain number of lattice sites $\mu\in\mathbb{Z}$. To account for this, if we wish to consider Gaussian state in the present framework with a non-zero `mean,' we must include symmetric positive and negative orientation modes. (This value $\mu$ is not the mean of $\hat{V}$ but rather of $|\hat{V}|$, see Appendix~\ref{app:absolute-value} for details) Therefore, a generic Gaussian state `centered' around a point $\mu\in\mathbb{Z}$ will be of the form

\begin{equation}\label{eq:gaussian-states-general}
    |\psi\rangle=\frac{c}{\sqrt{2}}\sum_{n\in\mathbb{Z}} \left[e^{-(nl-\mu l)^2/2\sigma^2}+e^{-(nl+\mu l)^2/2\sigma^2}\right]|nl\rangle.
\end{equation}

This may equivalently be written as

\begin{equation}
    |\psi\rangle=\frac{c}{\sqrt{2}}\sum_{n\in\mathbb{Z}} e^{-(nl)^2/2\sigma^2}\left(|nl+\mu l\rangle+|nl-\mu l\rangle\right).
\end{equation}

The normalization in this case is then given by

\begin{equation}
    |c|^2=\frac{1}{\sum_{n\in\mathbb{Z}}\left[e^{-(nl-\mu l)^2/2\sigma^2}+e^{-(nl+\mu l)^2/2\sigma^2}\right]^2}
\end{equation}

\noindent which reduces to~\eqref{eq:norm} when $\mu=0$. Since a choice of units is irrelevant to the physics of the theory, we shall henceforth set $l=l_P=1$. Expanding in $u:=\exp\{-1/\sigma^2\}$, one sees that

\begin{equation}\label{eq:norm-general}
    |c|^2=\frac{1}{2\left(1+e^{-(\mu/\sigma)^2}\right)\vartheta_3(u)}
\end{equation}

\noindent where $\vartheta_3(u)=\vartheta_3(0,u)$ is the third Jacobi theta function defined by

\begin{equation*}
    \vartheta_3(z,q):=\sum_{n\in\mathbb{Z}}q^{n^2}e^{2niz}.
\end{equation*}

This function is well-defined for all $u\in[0,1]$ and so for all values of $\sigma\neq 0$. One may readily evaluate the following limits:

\begin{equation}\label{eq:norm-asym}
    \lim_{\sigma\to 0}|c|^2=0,\qquad\lim_{\sigma\to\infty}|c|^2=\frac{1}{2}.
\end{equation}

Monotonicity ensures that $0<|c|^2<1/2$ for all values of $\sigma>0$ and $\mu$. We could just as well define Gaussian states in the dual $\beta$-representation (i.e. holonomy-flux coordinates in $L^2(\mathbb{R}_{\text{Bohr}},d\mu)$) as done in~\cite{corichi:2012,corichi:2011,velhinho:2007}, however, the volume representation is easier to calculate in.

\section{Exact Gaussian Fluctuations}
\label{sec:gaussian-fluctuations}
We now compute the exact solutions for the fluctuations of $\hat{V}$ and $\frac{1}{\lambda}\hat{S}_\lambda$ for Gaussian states in the volume representation. For a given $l$, the super-selection sector of the non-separable polymer Hilbert space is the subspace generated by the lattice of volume eigenstates of the form $|nl\rangle$ for $n\in\mathbb{Z}$. Fixing $l=1$ as we have done makes this space unique. Such a sector is isomorphic to the usual separable Hilbert space $\ell^2$, and so if all of the relevant operators of the theory are closed on such a sector, we may simply discuss that particular sector, whence the theory reduces to a quantum theory on a separable Hilbert space.

If one transforms away the the factors of $V_0$ in the canonical variables of the theory before quantization (e.g.~\cite{ashtekar:2003}), there is no such factor in the holonomy operator, and so one may choose the value of $\lambda$ to be a scaled integer multiple of $l$, reducing LQC to a separable Hilbert space. However, if one keeps the original coordinates, as we have done, the holonomy operator is not closed on the superselection sector unless $\lambda$ is made $V_0$-dependent. This detail will be important in the following analysis.

It is easy to show, due to parity symmetry, that $\langle\hat{V}\rangle=0$ for Gaussian states. Following Willis~\cite{willis:2004}, we may compute $\langle\hat{V}^2\rangle$  for Gaussian states using the Poisson summation formula. Given a function $g(y)$, this formula states that

\begin{equation*}
    \sum_{n\in\mathbb{Z}}g(x+n)=\sum_{n\in\mathbb{Z}}e^{i2\pi xn}\int_{-\infty}^{\infty}g(y) e^{-i2\pi yn}dy.
\end{equation*}

For Gaussian states, we then have

\begin{gather*}
    \langle\hat{V}^2\rangle=\sum_{n\in\mathbb{Z}}g(0+n)\\ g(y):=\frac{|c|^2}{2}y^2\left(e^{-(y-\mu)^2/\sigma^2}+e^{-(y+\mu)^2/\sigma^2}+2e^{-(y^2+\mu^2)/\sigma^2}\right)
\end{gather*}

We may evaluate this expression (noting that $\mu\in\mathbb{Z}$, whence the complex phase which arises from integration vanishes) to obtain

\begin{figure}[h!]
\centering
\includegraphics[width=0.4\textwidth]{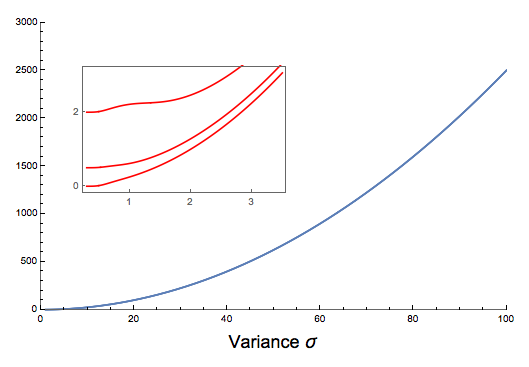}
\caption{The magnitude of the fluctuations $\langle\hat{V}^2\rangle$ for Gaussian states with variance $\sigma$ about $\alpha\lambda/V_0=1$ with $\mu=0,1,2$ (with minima increasing with $\mu$).}
\label{fig:volume-fluct}
\end{figure}

\begin{widetext}
\begin{equation}\label{eq:volume-moments}
    \begin{split}
        \langle\hat{V}^2\rangle=&\sum_{n\in\mathbb{Z}}\int_{-\infty}^{\infty}y^2\left(e^{-(y-\mu)^2/\sigma^2}+e^{-(y+\mu)^2/\sigma^2}+2e^{-(y^2+\mu^2)/\sigma^2}\right)e^{-i2\pi yn}dy\\
        =&\frac{|c|^2\sqrt{\pi}}{2}\sigma\sum_{n\in\mathbb{Z}}e^{-(\pi\sigma n)^2}\left[2\mu^2+\sigma^2(1+e^{-\mu^2/\sigma^2})(1-2\pi^2n^2)\right]\\
        =&\frac{|c|^2\sqrt{\pi}}{2}\sigma\left\{\left[2\mu^2+\sigma^2\left(1+e^{-\mu^2/\sigma^2}\right)\right]\vartheta_3(v)-2\pi^2\sigma^4e^{-\pi^2\sigma^2}\left(1+e^{-\mu^2/\sigma^2}\right)\left(\frac{d}{dv}\vartheta_3(v)\right)\right\}\\
        =&\frac{\sqrt{-\ln v}}{4\sqrt{\pi}\left(1+e^{\mu^2\pi^2/\ln v}\right)\vartheta_3(v^{1/\pi^2\sigma^4})}\left\{\left[2\mu^2-\frac{\ln v}{\pi^2}\left(1+e^{\mu^2\pi^2/\ln v}\right)\right]\vartheta_3(v)-\left(1+e^{\mu^2\pi^2/\ln v}\right)\left(\frac{2v(\ln v)^2}{\pi^2}\right)\left(\frac{d}{dv}\vartheta_3(v)\right)\right\}
    \end{split}
\end{equation}
\end{widetext}

\noindent where $v:=\exp\{-\pi^2\sigma^2\}=u^{\pi^2\sigma^4}$. Note that this is independent of $\lambda$ and $V_0$. Since $\langle\hat{V}\rangle=0$, we see that $\Delta^2_{\hat{V}}=\langle\hat{V}^2\rangle$. These volume fluctuations are computed numerically and plotted against $\sigma$ in Figure~\ref{fig:volume-fluct}.

We may now look at the fluctuations in the conjugate operator $\frac{1}{\lambda}\hat{S}_\lambda$. First, it is known that parity transformations take $\hat{S}_\lambda\mapsto -\hat{S}_\lambda$ (cf. the Appendix in~\cite{ashtekar:2009}). Thus, for the states in question, we readily see that $\langle\frac{1}{\lambda}\hat{S}_\lambda\rangle=0$. Finally, we have

\begin{equation}\label{eq:sin-squared-expectation}
\begin{split}
    \left\langle\left(\frac{1}{\lambda}\hat{S}_\lambda\right)^2\right\rangle=&\left\langle \left(-\frac{\hat{h}_\lambda-\hat{h}^*_\lambda}{2i\lambda}\right)^2\right\rangle\\
    =&\frac{1}{4\lambda^2}\left\langle 2\hat{I}-\hat{h}^2_\lambda-\hat{h}^{*2}_\lambda\right\rangle\\
    =&\frac{1}{4\lambda^2}\left\langle 2\hat{I}-\hat{h}_{2\lambda}-\hat{h}_{-2\lambda}\right\rangle.
\end{split}
\end{equation}

For the Gaussian state about $\mu$, we have

\begin{widetext}
\begin{equation}\label{eq:shift-squared}
\begin{split}
    \langle\hat{h}_{\pm 2\lambda}\rangle=&\frac{|c|^2}{2}\sum_{m,n\in\mathbb{Z}}e^{-m^2/2\sigma^2}e^{-n^2/2\sigma^2}\left(\langle m+\mu|+\langle m-\mu|\right)\left(\bigg|n+\mu\pm\frac{2\alpha\lambda}{V_0}\bigg\rangle+\bigg|n-\mu\pm\frac{2\alpha\lambda}{V_0}\bigg\rangle\right)\\
    =&\frac{|c|^2}{2}\sum_{m,n\in\mathbb{Z}}e^{-m^2/2\sigma^2}e^{-n^2/2\sigma^2}\bigg[2\delta\left(m-n\mp\frac{2\alpha\lambda}{V_0}\right)+\delta\left(m-n+2\mu\mp\frac{2\alpha\lambda}{V_0}\right)+\delta\left(m-n-2\mu\mp\frac{2\alpha\lambda}{V_0}\right)\bigg]\\
    =&\begin{cases}\frac{|c|^2}{2}\bigg[2u^{2(\pm\alpha\lambda/V_0)^2}\left(\sum_{n\in\mathbb{Z}}u^{n^2}u^{\pm2n\alpha\lambda/V_0}\right)+u^{2(\pm\alpha\lambda/V_0-\mu)^2}\left(\sum_{n\in\mathbb{Z}}u^{n^2}u^{2n(\pm\alpha\lambda/V_0-\mu)}\right),& \frac{2\alpha\lambda}{V_0}\in\mathbb{Z}
    \\\qquad+u^{2(\pm\alpha\lambda/V_0+\mu)^2}\left(\sum_{n\in\mathbb{Z}}u^{n^2}u^{2n(\pm\alpha\lambda/V_0+\mu)}\right)\bigg]&\\
    0&\text{otherwise}\end{cases}
\end{split}
\end{equation}
\end{widetext}

\noindent where $u=\exp\{-1/\sigma^2\}$. The last line is obtained by noting from Cauchy's criterion that the series converges for all values of the relevant parameters. Thus, we may expand the sum and evaluate each Kronecker-$\delta$ term independently. We see the dichotomy in cases here because, when $\frac{2\alpha\lambda}{V_0}\notin\mathbb{Z}$, the action of $\hat{h}_{2\lambda}$ shifts every term of the Gaussian state off of the permitted lattice sites; it is not closed on a superselection sector. In terms of operator-algebraic considerations, this dichotomy is related to rotation algebras, see Appendix~\ref{app:rotation-algebras}. The inner-product is unforgiving here; since none of the offset terms line up with any of the other points on the lattice, the state overlap vanishes everywhere. We may prevent this from happening by choosing $\lambda$ to be an integer multiple of $V_0/2\alpha$. Let us proceed by considering both cases.

\subsection{The non-integer case}

Suppose that $\frac{2\alpha\lambda}{V_0}\notin\mathbb{Z}$. Then we may readily compute from~\eqref{eq:sin-squared-expectation} that

\begin{equation*}
    \left\langle\left(\frac{1}{\lambda}\hat{S}_\lambda\right)^2\right\rangle=\frac{1}{2\lambda^2}.
\end{equation*}

Thus, since $\langle\hat{V}\rangle=\langle\frac{1}{\lambda}\hat{S}_\lambda\rangle=0$, we see that

\begin{widetext}
\begin{equation}
\begin{split}
    \Delta_V^2\Delta_{\frac{1}{\lambda}S_\lambda}^2=&\left\langle\hat{V}^2\right\rangle\left\langle\left(\frac{1}{\lambda}\hat{S}_\lambda\right)^2\right\rangle\\
    =&\frac{\sqrt{-\ln v}}{8\lambda^2\sqrt{\pi}\left(1+e^{\mu^2\pi^2/\ln v}\right)\vartheta_3(v^{1/\pi^2\sigma^4})}\left\{\left[2\mu^2-\frac{\ln v}{\pi^2}\left(1+e^{\mu^2\pi^2/\ln v}\right)\right]\vartheta_3(v)-\left(1+e^{\mu^2\pi^2/\ln v}\right)\left(\frac{2v(\ln v)^2}{\pi^2}\right)\left(\frac{d}{dv}\vartheta_3(v)\right)\right\}
\end{split}
\end{equation}
\end{widetext}

\noindent with $v=\exp\{-\pi^2\sigma^2\}$. Overall, the fluctuations here behave just like $\langle\hat{V}^2\rangle$ in Figure~\ref{fig:volume-fluct} with an overall suppression by a factor $2\lambda^2$. If we allow $\lambda$ to be an arbitrary independent parameter of the theory, this case is generic. Let us now consider the integer case.

\subsection{The integer case}

Suppose $\frac{2\alpha\lambda}{V_0}\in\mathbb{Z}$. From~\eqref{eq:shift-squared}, taking $q=u=\exp\{-1/\sigma^2\}$ and making a perspicuous choice of $z$ for each term as the arguments for the Jacobi $\vartheta$-function $\vartheta_3(z,q)$, we obtain

\begin{widetext}
\begin{equation}
    \begin{split}
     \langle\hat{h}_{\pm 2\lambda}\rangle=&\frac{|c|^2}{2}\bigg[2u^{2(\pm\alpha\lambda/V_0)^2}\vartheta_3\left(\frac{\mp\alpha\lambda/V_0}{i\sigma^2},u\right)+u^{2(\pm\alpha\lambda/V_0-\mu)^2}\vartheta_3\left(\frac{\mp\alpha\lambda/V_0+\mu}{i\sigma^2},u\right)+u^{2(\pm\alpha\lambda/V_0+\mu)^2}\vartheta_3\left(\frac{\mp\alpha\lambda/V_0-\mu}{i\sigma^2},u\right)\bigg].
    \end{split}
\end{equation}
\end{widetext}

\begin{figure}[h!]
\centering
\includegraphics[width=0.4\textwidth]{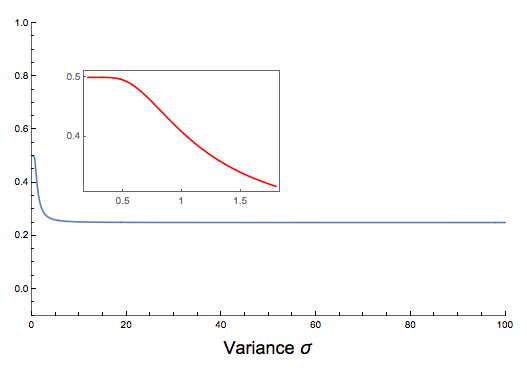}
\caption{The magnitude of the fluctuations $\langle\left(\frac{1}{\lambda}\hat{S}_\lambda\right)^2\rangle$ for Gaussian states with variance $\sigma$ and volume mean $\mu=0$.}
\label{fig:sine-fluct}
\end{figure}

This calculation may be substituted into~\eqref{eq:sin-squared-expectation} to obtain an expression for the expected variance in $\frac{1}{\lambda}\hat{S}_\lambda$. This expectation value is plotted against $\sigma$ in Figure~\ref{fig:sine-fluct}. One can see readily that, for large $\sigma$, these fluctuations become essentially constant at $1/4$.

\begin{figure}[h!]
\centering
\includegraphics[width=0.4\textwidth]{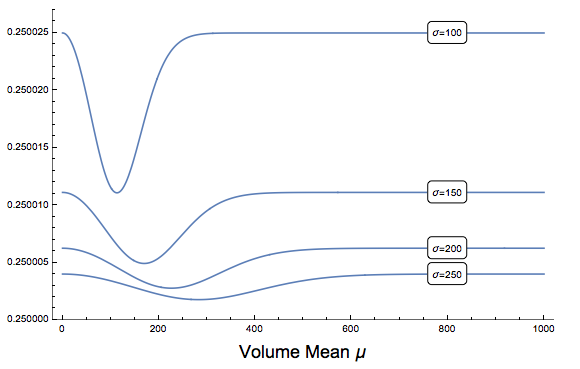}
\caption{The magnitude of the fluctuations $\langle\left(\frac{1}{\lambda}\hat{S}_\lambda\right)^2\rangle$ for Gaussian states with variance $\sigma$ and mean $\mu$.}
\label{fig:sine-fluct-mean}
\end{figure}

We may also check the behaviour of these fluctuations is $\mu$ varies. This is illustrated in Figure~\ref{fig:sine-fluct-mean}. Note that this quantity rapidly converges as $\mu$ gets large for any fixed value of $\sigma$ with an asymptotic minimum at $1/4$ as $\sigma\to\infty$ and $\mu\to\infty$.

\begin{figure}[h!]
\centering
\includegraphics[width=0.4\textwidth]{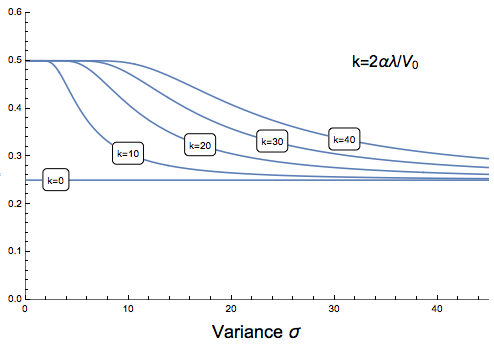}
\caption{The magnitude of the fluctuations $\langle\left(\frac{1}{\lambda}\hat{S}_\lambda\right)^2\rangle$ for Gaussian states with variance $\sigma$ at different fixed integer values of $\alpha\lambda/V_0$ (with $\mu=0$).}
\label{fig:sine-fluct-lambda}
\end{figure}

In the integer case, the sine fluctuations also vary as $\alpha\lambda/V_0$ changes. We plot these fluctuations as a function of $\sigma$ for several integer values of $2\alpha\lambda/V_0$ in Figure~\ref{fig:sine-fluct-lambda}. We see that the rapid convergence to 1/4 persists, but larger integer values of $\alpha\lambda/V_0$ result in a slower initial rate of convergence.

We may now analyze the overall fluctuations $\Delta_V^2\Delta_{\frac{1}{\lambda}S_\lambda}^2$. By supposing $\frac{2\alpha\lambda}{V_0}\in\mathbb{Z}$, we assume that there exists some integer $k$ such that $\lambda=kV_0/2\alpha$. Thus, substituting this value, from~\eqref{eq:volume-moments} and~\eqref{eq:sin-squared-expectation}, the overall expression for the fluctuations in the integer case is:

\begin{widetext}
\begin{equation}\label{eq:full-fluctuations-int}
    \begin{split}
        \Delta_V^2\Delta_{\frac{1}{\lambda}S_\lambda}^2=&\left\langle\hat{V}^2\right\rangle\left\langle\left(\frac{1}{\lambda}\hat{S}_\lambda\right)^2\right\rangle\\
        =&\frac{\alpha^2\sqrt{\pi}\sigma}{4k^2V_0^2\left(1+e^{-\mu^2/\sigma^2}\right)\vartheta_3(e^{-1/\sigma^2})}\left\{\left[2\mu^2+\sigma^2\left(1+e^{-\mu^2/\sigma^2}\right)\right]\vartheta_3(e^{-\pi^2\sigma^2})-2\pi^2\sigma^4e^{-\pi^2\sigma^2}\left(1+e^{-\mu^2/\sigma^2}\right)\left(\frac{d}{dv}\vartheta_3(v)\right)\right\}\\
        &\times\bigg(2-\frac{1}{4(1+e^{-\mu^2/\sigma^2})\vartheta_3\left(e^{-1/\sigma^2}\right)}\bigg\{2e^{-2(\alpha\lambda/V_0)^2/\sigma^2}\bigg[\vartheta_3\left(\frac{-\alpha\lambda/V_0}{i\sigma^2},e^{-1/\sigma^2}\right)+\vartheta_3\left(\frac{\alpha\lambda/V_0}{i\sigma^2},e^{-1/\sigma^2}\right)\bigg]\\
        &+e^{-2(\alpha\lambda/V_0-\mu)^2/\sigma^2}\bigg[\vartheta_3\left(\frac{-\alpha\lambda/V_0+\mu}{i\sigma^2},e^{-1/\sigma^2}\right)+\vartheta_3\left(\frac{\alpha\lambda/V_0-\mu}{i\sigma^2},e^{-1/\sigma^2}\right)\bigg]\\
        &+e^{-2(\alpha\lambda/V_0+\mu)^2/\sigma^2}\bigg[\vartheta_3\left(\frac{-\alpha\lambda/V_0-\mu}{i\sigma^2},e^{-1/\sigma^2}\right)+\vartheta_3\left(\frac{\alpha\lambda/V_0+\mu}{i\sigma^2},e^{-1/\sigma^2}\right)\bigg]\bigg\}\bigg)
    \end{split}
\end{equation}
\end{widetext}

\noindent with $v=\exp\{-\pi^2\sigma^2\}$ in the $\frac{d}{dv}\vartheta_3(v)$ term for brevity (where all other $v$ and $u$ terms have been expanded in~$\sigma$).

We may now state the two main results of this section. First, if we analyze the numerically generated plots for the integer case in Figures~\ref{fig:volume-fluct},~\ref{fig:sine-fluct},~\ref{fig:sine-fluct-mean}, and~\ref{fig:sine-fluct-lambda}, we see that these fluctuations diverge as $\sigma$ gets large, effectively scaling as $\langle\hat{V}^2\rangle/4$. This is very different from what is observed in ordinary quantum mechanics where the fluctuations of Gaussian states in their canonical coordinates are not only bounded, but constant for all values of $\sigma$.

The other important realization from this discussion is that from~\eqref{eq:full-fluctuations-int}, we observe

\begin{equation}
    \lim_{V_0\to\infty}\Delta_V^2\Delta_{\frac{1}{\lambda}S_\lambda}^2=0.
\end{equation}

This was the claim posited by Rovelli and Wilson-Ewing in~\cite{rovelli:2014}. This result was, for them, somewhat unexpected since $V_0$ is just a regularization parameter and thus should not be physically significant. However, their original analysis was non-constructive; they examined the lower bound of~\eqref{eq:Heis-Uncertainty} but did not construct explicit semi-classical states for which the fluctuations saturate this bound in the appropriate limits. Here, we have provided such explicit states and derived this result exactly. The benefit to this is that it allows us to see where this surprising result comes from.

This scaling of fluctuations with $V_0$ arises only when one introduces a $V_0$-dependence to $\lambda$. One cannot take such a transformation to just be a scaling of the theory variables because, if $\lambda$ is fixed, the shift operator $\hat{h}_{\lambda}$ will discontinuously jump between being closed on the superselection sector, and not being closed. Hence, this dependency is connected to the basic Hilbert space structure of the theory, and so it is unsurprising that it should impact the resulting phenomenology. Interestingly, however, while scaling $V_0$ allows one to reduce the fluctuations of these Gaussian states, it does not ensure that fluctuations are minimal for any fixed value of $V_0$. We shall show this in the next section by comparing the results presented here with the lower bound of the Robertson-Schr\"odinger inequality.

It should be noted that if one generalizes this analysis to the context of spatially compact spacetimes with volume $V_{\text{Max}}$, then the regularization parameter $V_0$ only makes sense when $V_0\leq V_{\text{Max}}$. Indeed, it is most natural to simply fix $V_0=V_{\text{Max}}$. As such, the $V_0\to\infty$ limit is no longer possible in this context; one may only scale the fiducial volume so far. This means that there is a limit to how far one may suppress the fluctuations of Gaussian states using this procedure, and so there will still generally be many Gaussian states with extremely large fluctuations.

\section{Saturating Uncertainty}
\label{sec:saturation}
We now examine the lower bound of the uncertainty inequality~\eqref{eq:RS-volume-flux} (which reduces to~\eqref{eq:Heis-Uncertainty}) for the fluctuations of Gaussian states. Generically, the lower bound is given by

\begin{equation*}
    \left(\frac{\alpha}{2V_0}\right)^2|\langle\hat{C}_\lambda\rangle|^2=\left(\frac{\alpha}{4V_0}\right)^2|\langle\hat{h}_{\lambda}+h_{-\lambda}\rangle|^2.
\end{equation*}

Using an identical sector closure argument as above, $\langle\hat{h}_{\pm\lambda}\rangle=0$ if $\alpha\lambda/V_0\not\in\mathbb{Z}$, whence the lower bound is zero. That is, we now have a split between the cases when  $k=2\alpha\lambda/V_0$ is a is an even integer or not. In case where $k$ is not an even integer, this lower bound is zero. Otherwise, substituting this value for $\lambda$, the lower bound may be computed to be

\begin{widetext}
\begin{equation}\label{eq:RS-Gaussian-LB}
\begin{split}
    \left(\frac{\alpha}{2V_0}\right)^2|\langle\hat{C}_\lambda\rangle|^2=&\left(\frac{\alpha}{8V_0\left(1+e^{-(\mu/\sigma)^2}\right)\vartheta_3(e^{-1/\sigma^2})}\right)^2\\
    &\times\bigg|2e^{-k^2/8\sigma^2}\vartheta_3\left(\frac{-k}{i4\sigma^2},e^{-1/\sigma^2}\right)+e^{-2(k/4-\mu)^2/\sigma^2}\vartheta_3\left(\frac{-k/4+\mu}{i\sigma^2},e^{-1/\sigma^2}\right)\\
    &+e^{-2(k/4+\mu)^2/\sigma^2}\vartheta_3\left(\frac{-k/4-\mu}{i\sigma^2},e^{-1/\sigma^2}\right)+2e^{-k^2/8\sigma^2}\vartheta_3\left(\frac{k}{i4\sigma^2},e^{-1/\sigma^2}\right)\\
    &+e^{-2(-k/4-\mu)^2/\sigma^2}\vartheta_3\left(\frac{k/4+\mu}{i\sigma^2},e^{-1/\sigma^2}\right)+e^{-2(-k/4+\mu)^2/\sigma^2}\vartheta_3\left(\frac{k/4-\mu}{i\sigma^2},e^{-1/\sigma^2}\right)\bigg|^2.
\end{split}
\end{equation}
\end{widetext}

The degree to which this bound is saturated by the actual fluctuations is given by the difference between~\eqref{eq:full-fluctuations-int} and~\eqref{eq:RS-Gaussian-LB}. There are three cases to consider; the non-integer case where $\alpha\lambda/V_0\not\in\mathbb{Z}$ whence both sides of the inequality are simplified, the odd-$k$ case where $2\alpha\lambda/V_0\in\mathbb{Z}$ but $\alpha\lambda/V_0\not\in\mathbb{Z}$, whence the exact value is complicated but the lower bound vanishes, and the even-$k$ case where $\alpha\lambda/V_0\in\mathbb{Z}$, whence both sides have a complicated form.

\subsection{Non-integer case}

When $2\alpha\lambda/V_0\not\in\mathbb{Z}$, the Roberson-Schr\"odinger-inequality becomes

\begin{widetext}
\begin{equation}\label{eq:saturating-bound-NHI}
\begin{split}
    0\leq&\frac{\sqrt{\pi}\sigma}{8\lambda^2\left(1+e^{-\mu^2/\sigma^2}\right)\vartheta_3(e^{-1/\sigma^2})}\left\{\left[2\mu^2+\sigma^2\left(1+e^{-\mu^2/\sigma^2}\right)\right]\vartheta_3(v)-2\pi^2\sigma^4e^{-\pi^2\sigma^2}\left(1+e^{-\mu^2/\sigma^2}\right)\left(\frac{d}{dv}\vartheta_3(v)\right)\right\}.
\end{split}
\end{equation}
\end{widetext}

Any roots of this expression correspond to all of the instances in which the Robertson-Schr\"odinger inequality is saturated. However, by inspecting the behaviour of $\langle\hat{V}^2\rangle$ in Figure~\ref{fig:volume-fluct}, we see that it has no roots. The next problem, then, is to determine under what circumstances these fluctuations are minimal, if non-zero. But the right-hand side of the inequality is monotonically increasing so we may conclude that it always decreases as $\sigma\to 0$ and is thus never minimal. Every Gaussian state in this case always violates the Robertson-Schr\"odinger inequality, and is only made optimal by taking $\sigma$ to be as small as possible.

\subsection{Odd-$k$ case}

When $k=2\alpha\lambda/V_0$ is an odd integer, the exact fluctuations are given by~\eqref{eq:full-fluctuations-int}, while the lower bound is zero. Thus, the analysis is identical to the above so we exclude it for brevity, noting that the uncertainty inequality is never saturated, and the fluctuations asymptotically diverge from the lower bound.

\subsection{Even-$k$ case}

\begin{figure}[h!]
\centering
\includegraphics[width=0.4\textwidth]{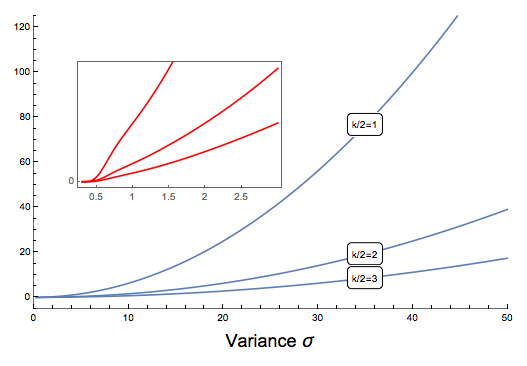}
\caption{The difference between the exact fluctuations $\Delta_{\hat{V}}^2\Delta_{\frac{1}{\lambda}\hat{S}_\lambda}^2$ and the lower bound given by the Robertson-Schr\"odinger inequality for a Gaussian state with variance $\sigma$ and mean $\mu=0$ in the integer case $\alpha\lambda/V_0\in\mathbb{Z}$ (where $\lambda=\frac{kV_0}{2\lambda}$ for $k\in\mathbb{Z}$).}
\label{fig:fluct-diff-k}
\end{figure}

Now suppose that $\alpha\lambda/V_0\in\mathbb{Z}$ so that $k$ is an even integer. In this case, the uncertainty relation is given by ~\eqref{eq:full-fluctuations-int} being greater than or equal to~\eqref{eq:RS-Gaussian-LB}. One can readily check that, since $e^{-1/\sigma^2}<1$ and $e^{-\pi^2\sigma^2}<1$ for all values of $\sigma> 0$, all of the $\vartheta$-terms in~\eqref{eq:full-fluctuations-int} and~\eqref{eq:RS-Gaussian-LB} are suppressed by their exponential prefactors exponentially quickly as $k$ gets large. Thus, in the large $k$ regime, the fluctuations in the even-$k$ case reduce to the non-integer case. We therefore see that, for large holonomy lengths $\lambda$, that is the generic case. We plot the difference between the exact fluctuations and the lower bound for several values of $k$ in Figure~\ref{fig:fluct-diff-k}. Again, these fluctuations do not saturate the uncertainty relation.

The important fact from this section is that, contrary to the case in ordinary quantum mechanics, Gaussian states do \textit{not} saturate uncertainty relations in LQC.

\section{Squeezed States}
\label{sec:squeezed-states}
We have seen that Gaussian states do not minimize the uncertainty relation exactly, though may do so asymptotically in the $V_0\to\infty$ limit. This naturally raises the question: are there \textit{any} well-defined states on the superselection sector which minimize uncertainty? In this section, I sketch an argument that, barring the trivial zero-volume state $|0\rangle$, the answer is likely no. This argument is not a formal proof but rather of a rough motivation.

For any symmetric operators $\hat{A}$ and $\hat{B}$, it is known~\cite[p. 244]{hall:2013} that~\eqref{eq:Heis-Uncertainty} is saturated by a pure state $|\psi\rangle$ if and only if $|\psi\rangle$ is an eigenstate of $\hat{A}$ or $\hat{B}$, or if $|\psi\rangle$ is an eigenstate of $\hat{A}+i\xi\hat{B}$ for some $\xi\in\mathbb{R}$. The last case are the so-called $\xi$-squeezed coherent states of $\hat{A}$ and $\hat{B}$.

As a simplifying assumption, let us suppose that $\alpha\lambda/V_0=1$ so that $\hat{h}_\lambda|V\rangle=|V+1\rangle$. Taking more general integer values will in principle yield a similar result, but the necessary analysis becomes too unwieldy to be presented here. Let us suppose for reductio that $|\psi\rangle$ is a state on the superselection lattice which is an eigenstate of $\hat{V}+\frac{i\xi}{\lambda}\hat{S}_\lambda$ and thus minimizes the uncertainty relation for the operators $\hat{V}$ and $\frac{1}{\lambda}\hat{S}_\lambda$. Then $|\psi\rangle$ may be written as

\begin{equation}
    |\psi\rangle=\sum_{n\in\mathbb{Z}}\psi(n)|n\rangle
\end{equation}

\noindent where $\psi(n):\mathbb{Z}\to\mathbb{C}$ obeys $\sum_n|\psi(n)|^2<\infty$. Then we have for some eigenvalue $A$:

\begin{widetext}
\begin{equation}
\begin{split}
    A|\psi\rangle=\left(\hat{V}+\frac{i\xi}{\lambda}\hat{S}_\lambda\right)|\psi\rangle=&\sum_{n\in\mathbb{Z}}n\psi(n)|n\rangle-\psi(n)\frac{\xi}{2\lambda}\left(|n-1\rangle-|n+1\rangle\right)\\
    =&\sum_{n\in\mathbb{Z}}\left(n\psi(n)-\frac{\xi}{2\lambda}\psi(n+1)+\frac{\xi}{2\lambda}\psi(n-1)\right)|n\rangle=\sum_{n\in\mathbb{Z}}A\psi(n)|n\rangle.
\end{split}
\end{equation}
\end{widetext}

The coefficients $\psi(n)$ only satisfy this condition if they obey the difference equation

\begin{equation}
    2(A-n)\psi(n)=\frac{\xi}{\lambda}[\psi(n-1)-\psi(n+1)]
\end{equation}

\noindent for all $n\in\mathbb{Z}$. This difference equation may be solved to yield:

\begin{equation}
    \psi(n)=c_1I_{n-A}\left(\frac{-\xi}{\lambda}\right)+c_2K_{n-A}\left(\frac{\xi}{\lambda}\right)
\end{equation}

\noindent for arbitrary constants $c_1$ and $c_2$ where $I$ and $K$ are the modified Bessel functions of the first and second kind, respectively. In order for a state with coefficients $\psi(n)$ to be an element of the Hilbert space, we require that

\begin{equation}
    \sum_{n\in\mathbb{Z}}\left|c_1I_{n-A}\left(\frac{\xi}{\lambda}\right)+c_2K_{n-A}\left(\frac{-\xi}{\lambda}\right)\right|^2<\infty.
\end{equation}

However, for any fixed $\xi$, as $n\to+\infty$, $K_{n-A}(\frac{\xi}{\lambda})$ diverges, yet $I_{n-A}(\frac{-\xi}{\lambda})$ converges to zero (and so cannot be scaled to counter-act this divergence) so for this condition to be satisfied, we must set $c_2=0$. Thus, the only viable states are those of the form

\begin{equation}
    \psi(n)=c_1I_{n-\lambda}\left(\frac{-\xi}{\lambda}\right)
\end{equation}

However, as $n\to-\infty$, one may check that $|I_{n-\lambda}(\frac{-\xi}{\lambda})|^2$ likewise diverges. Thus, the boundedness condition further requires that $c_1=0$ as well. Hence, there are no non-zero vectors in this Hilbert space which are $\xi$-squeezed states for any value of $\xi$. Thus, there are no $\xi$-squeeze states.

The only alternatives, then, for minimizing uncertainty are eigenstates of $\hat{V}$ or $\frac{1}{\lambda}\hat{S}_\lambda$. The only eigenstate of $\hat{V}$ which respects the requisite parity symmetry $\psi(V)=\psi(-V)$ is the zero-volume state $|\psi\rangle=|0\rangle$. Thus, this is the trivial unique volume eigenstate which minimizes uncertainty. It is non-trivial to compute eigenstates of $\frac{1}{\lambda}\hat{S}_\lambda$, but it is conjectured that any such states will likewise be unphysical, or at least fail to be sharply peaked, and so fail to be semi-classical.

\section{Conclusion}
Volume-regularized loop quantum cosmology is not a single theory, but rather a large class of different theories differentiated from one another by their associated choice of fiducial volume $V_0$ and their chosen holonomy length $\lambda$. We have here constructed generic families of Gaussian states on superselection sectors in these theories and computed their fluctuations with respect to the canonical observables of these theories, namely, volume, and the sine of holonomy. These fluctuations were then compared with their fundamental lower bound given by the Robertson-Schr\"odinger inequality, a generalization of the uncertainty principle. Three salient results were shown.

\begin{enumerate}
    \item For a Gaussian state with a fixed width $\sigma$ and a fixed parity-symmetric mean $\mu$, one can always choose a theory of LQC with a sufficiently large fiducial volume $V_0$ such that the fluctuations of this state become negligible. This fails to be true for spatially compact spacetimes.
    \item Within a fixed theory of LQC (with a given finite $V_0$), one can always find a Gaussian state with sufficiently large $\sigma$ such that its fluctuations become arbitrarily large.
    \item The relation between $\lambda$ and $V_0$ play an important role in determining the phenomenology of a theory of LQC, namely, they determine whether or not the holonomy operator $\hat{h}_\lambda$ is closed on the relevant superselection sector.
\end{enumerate}

These results indicate that Gaussian states are not universally semi-classical in theories of LQC and that the semi-classical sector of a theory of LQC depends sensitively upon its basic constitutive parameters.

\appendix

\section{The $|\hat{V}|$ Operator}
\label{app:absolute-value}
The `physical' volume of spacetime is independent of its orientation. Thus, the operator which corresponds to the `physical' volume of a quantum spacetime is not $\hat{V}$, but rather $|\hat{V}|$, defined on the polymer Hilbert space by

\begin{equation}
    |\hat{V}||V\rangle=\begin{cases}V|V\rangle,&V\geq 0\\-V|V\rangle,&V<0\end{cases}.
\end{equation}

In this view, the definition of a Gaussian state given in~\eqref{eq:gaussian-states-general} more closely resembles the traditional notion of a Gaussian state, for in this case, we see that $\langle|\hat{V}|\rangle\propto\mu$, and so $\mu$ is properly the mean. To see this, we may compute

\begin{widetext}
\begin{equation}
    |\hat{V}||\psi\rangle=\frac{c}{\sqrt{2}}\sum_{n>0}n\left(e^{-(n-\mu)^2/2\sigma^2}+e^{-(n+\mu)^2/2\sigma^2}\right)(|n\rangle+|-n\rangle)
\end{equation}
\end{widetext}

From which it follows that

\begin{equation}
        \langle|\hat{V}|\rangle=2|c|^2\sum_{n>0}n\left(e^{-(n-\mu)^2/\sigma^2}+e^{-(n+\mu)^2/\sigma^2}\right)^2
\end{equation}

\begin{figure}[h!]
\centering
\includegraphics[width=0.4\textwidth]{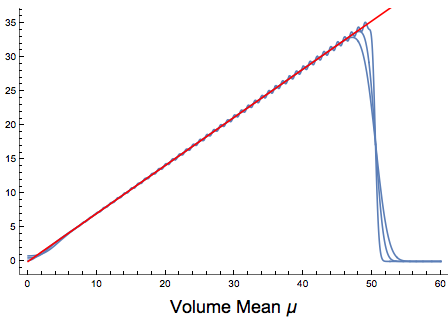}
\caption{The first 50 terms of the volume mean $\langle|\hat{V}|\rangle$ for Gaussian states with mean $\mu$ at values $\sigma=1,2,3$.}
\label{fig:abs-volume}
\end{figure}

This is hard to solve analytically, however, large $n$ terms get exponentially suppressed and so we may approximate this series for small $\mu$ and non-small $\sigma$ with a cutoff. Plotting the first 50 terms against $\mu$ with $\sigma=1,2,3$ in Figure~\ref{fig:abs-volume}, we see that this is nicely fit by $\mu/\sqrt{2}$ (the extra factor arises from the state normalization). We may readily note that $\hat{V}^2=|\hat{V}|^2$, and so the rest of the above analysis of oriented volume fluctuations persists in when orientation is dispenses with.

\section{$\hat{h}_\lambda$ Closure and Rotation Algebras}
\label{app:rotation-algebras}
In $C^*$-algebra theory, the rotation algebra $\mathcal{A}_\theta$ is characterized by the universal property of containing two unitary elements $U_1$ and $U_2$ which satisfy

\begin{equation}
    U_1U_2=e^{i2\pi\theta}U_2U_1.
\end{equation}

There are three cases of this algebra to consider: \textit{(i)} the trivial commutative case where $\theta\in\mathbb{Z}$, \textit{(ii)} the case where $\theta\in\mathbb{Q}$ (called a \textit{rational} rotation algebra), and \textit{(iii)} the case where $\theta\notin\mathbb{Q}$ (called an \textit{irrational} rotation algebra). The three cases are radically different, and the spectral theory of the later two (especially irrational rotation algebras) is notoriously rich.

In the present context, fixing $\lambda$, we see that the $C^*$-algebra generated by $\hat{V}$ and $\hat{h}_\lambda$ carries a representation of $\mathcal{A}_\theta$ for $\theta=\frac{2\alpha\lambda}{V_0}$. Specifically, defining a the unitary operator $\hat{U}=e^{i2\pi\hat{V}}$ generated by a series expansion in $\hat{V}$, we readily see that

\begin{equation}
    \hat{U}\hat{h}_\lambda=e^{i2\pi\theta}\hat{h}_\lambda\hat{U}.
\end{equation}

In the instance when $\frac{2\alpha\lambda}{V_0}\in\mathbb{Z}$ (whence $\hat{h}_\lambda$ is closed on the superselection sector), we see that $\mathcal{A}_\theta$ is commutative. In the case where $\hat{h}_\lambda$ is \textit{not} closed on the relevant superselection sector, one finds that $\mathcal{A}_\theta$ becomes a non-trivial rotation algebra and has much richer features.

\begin{acknowledgements}
I wish to thank Francesca Vidotto, Edward Wilson-Ewing, and Carlo Rovelli for valuable discussions and manuscript feedback. I acknowledge the support of the Natural Sciences and Engineering Research Council of Canada (NSERC), funding reference number USRA-554658-2020. This work is based on research supported by the John Templeton Foundation under grant \#61048.
\end{acknowledgements}

\bibliography{LQC-Fluctuations.bib}

\end{document}